\documentclass{PoS}

\usepackage{graphicx,multirow,subfig}
\usepackage{float}
\usepackage{bm}
\usepackage{amsmath}
\usepackage{amssymb}
\usepackage{amscd}
\usepackage{latexsym}
\usepackage{slashed}
\usepackage{color}
\usepackage{graphicx}
\usepackage{verbatim}

\newcommand{\be}{\begin{equation}}
\newcommand{\ee}{\end{equation}}

\title{A Composite 2-Higgs Doublet Model}

\ShortTitle{A Composite 2-Higgs Doublet Model}

\author{Stefania De Curtis\\
        INFN, Sezione di Firenze and Department of Physics and Astronomy, University of Florence,\\
        Via G. Sansone 1, 50019 Sesto Fiorentino, Italy \\
        E-mail: \email{decurtis@fi.infn.it}}

\author{\speaker{Luigi Delle Rose}\\
        INFN, Sezione di Firenze and Department of Physics and Astronomy, University of Florence,\\
        Via G. Sansone 1, 50019 Sesto Fiorentino, Italy \\
        E-mail: \email{luigi.dellerose@fi.infn.it}}

\author{Stefano Moretti\\
        School of Physics and Astronomy, University of Southampton,\\
        Highfield, Southampton SO17 1BJ, United Kingdom\\
        Particle Physics Department, Rutherford Appleton Laboratory, \\
        Chilton, Didcot, Oxon OX11 0QX, United Kingdom\\
        E-mail: \email{s.moretti@soton.ac.uk}}

\author{Key Yagyu\\
        Department of Physics, Osaka University, Toyonaka, Osaka 560-0043, Japan\\
        E-mail: \email{yagyu@het.phys.sci.osaka-u.ac.jp}}

\abstract{In the context of Composite Higgs Models we consider the realisation of an extended Higgs sector with two Higgs doublets arising as pseudo Nambu-Goldstone bosons from a $\textrm{SO}(6) \to \textrm{SO}(4) \times \textrm{SO}(2)$ breaking. 
The properties of the Higgses are obtained in terms of the fundamental parameters of the composite sector, such as masses, Yukawa and gauge couplings of the new spin-1/2 and spin-1 resonances. 
After computing the Higgs potential from the explicit breaking of the $\textrm{SO}(6)$ global symmetry by the partial compositeness of fermions and gauge bosons, the main focus is to
derive the phenomenological properties of the Higgs bosons and to highlight the main signatures of the Composite 2-Higgs Doublet Model at the Large Hadron Collider, including modifications to the SM-like Higgs couplings, production and decay channels of heavier Higgs states.}

\FullConference{
European Physical Society Conference on High Energy Physics - EPS-HEP2019 -\\
			10-17 July, 2019\\
			Ghent, Belgium}

\begin{document}

\section{The Composite 2-Higgs Doublet Model}
\label{sec:introduction}

Compositeness can naturally solve the hierarchy problem of the Standard Model (SM), in particular through the pseudo-Nambu-Goldstone  boson (pNGB) nature of the Higgs state. 
The idea is borrowed from QCD: the SM Higgs boson could be the analogue of the pion of the strong interactions. Just like there are mesons predicted by QCD, $\pi$, $\eta$, etc., there could be several Higgs states predicted by compositeness beyond the one discovered so far, as well as additional composite states (the equivalent of the $\rho$, $\omega$, etc. of QCD). 
In this respect, a natural setting  \cite{Mrazek:2011iu} is the Composite 2-Higgs Doublet Model (C2HDM) \cite{DeCurtis:2016scv,DeCurtis:2016tsm,DeCurtis:2017gzi}. 
It is built by enlarging the coset associate to the breaking of the global symmetry of the underlaying strong interactions to contain, beside the SM-like Higgs doublet, an additional one. The presence of an extra Higgs doublet is predicted in Supersymmetric (SUSY) models. We are here  proposing a composite realisation of such a scenario.
{
We focus on the Higgs sector of the C2HDM originating from the breaking ${\rm SO}(6)\to {\rm SO}(4)\times {\rm SO}(2)$ in presence of the {\sl partial compositeness} \cite{Kaplan:1991dc} driven by the third generation of the SM fermions.  
The masses of the Higgs bosons and their self-interactions are generated at one-loop level from the Coleman-Weinberg potential by the linear mixing between the (elementary) SM and the (composite) strong sector fields. As such, the masses and couplings are not free parameters, unlike in the elementary realisations, but depend upon the strong sector dynamics and present strong correlations among them which is one of the main features of Composite Higgs Models (CHMs).
A comparative study between the 2HDM arising from a composite dynamics and the one realised in SUSY models has been presented in \cite{DeCurtis:2018iqd}.
The scale of compositeness $f$ is typically within the energy reach of the LHC and the composite nature of the SM-like Higgs boson in the C2HDM can be assessed by exploiting the corrections of ${\cal O}(\xi)$ to its couplings, where $\xi=v_{\rm SM}^2/f^2$ with $v_{\rm SM}$ being the Vacuum Expectation Value (VEV) of the Higgs in the SM.
Since current (lower) limits on $f$ are of order 700-800 GeV, such deviations affect experimental observables only at the 5-10\% level making their observations a difficult task for the LHC.
For example, the ${\cal O}(\xi)$ corrections in the Yukawa interactions with top- and bottom-quarks or tau-leptons  are notoriously difficult to measure at the LHC as they are affected by a significant  QCD background. 
A much cleaner alternative is thus to probe the interactions of the SM-like Higgs with the gauge bosons, which are also affected by similar ${\cal O}(\xi)$ corrections. 
They are expected to be larger than the corresponding ones in elementary 2HDMs (E2HDM) \cite{Branco:2011iw} and more accessible at the High Luminosity LHC (HL-LHC).  
However, if these were to be found consistent with the E2HDM predictions, the last resort in the quest to disentangle the C2HDM from the E2HDM realisation would be to exploit the correlation among several observables of  the extra Higgs bosons.
Hence, under the above circumstances, it becomes mandatory to explore the scope of the HL-LHC and also of the High-Energy option of the LHC (HE-LHC) with 27 TeV centre of mass energy, a future project which is now under discussion. In particular, the HE-LHC would be necessary for processes involving the Higgs boson self-couplings which have rather small cross-sections  at the current LHC.
In the following we study under which CERN machine configurations one can access the processes $gg \to H \to hh \to b\bar b\gamma\gamma$ and $gg \to H\to t\bar t$ (in the semi-leptonic decay channel) which can be exploited to extract crucial features of the C2HDM, where $H$ and $h$ are the heaviest and the lightest (the SM-like one) of the two CP-even neutral Higgs states, respectively.
}

\section{Phenomenology}
\label{sec:results}
{
The construction of the model is described in \cite{DeCurtis:2018iqd,DeCurtis:2018zvh}. 
The fundamental parameters of the C2HDM correspond to the compositeness scale $f$, the gauge coupling of the underlaying strong interaction, the masses and Yukawas of the heavy top partners, as well as the mixing between the latter and the elementary top quark (which represents the leading contribution to the Coleman-Weinberg effective potential). 
The parameter space has been explored by scanning the compositeness scale in the range $(750, 3000)$ GeV and all the other mass parameters in the range $(-10,10)f$.
Phenomenologically acceptable configurations are obtained by requiring the vanishing of the two tree-level tadpoles of the CP-even Higgs bosons and the measured masses of the top quark and the SM-like Higgs boson. 
The masses of the heavier CP-even Higgs boson ($m_H$), the charged Higgs boson $(m_{H^\pm})$, the CP-odd Higgs boson ($m_A$), 
the mixing angle $\theta$ between the two CP-even states ($h,H$) and their couplings to fermions and bosons are all obtained from the fundamental parameters.
These quantities are then tested against experimental measurements through the HiggsBounds \cite{Bechtle:2013wla} and the HiggsSignals \cite{Bechtle:2013xfa} packages,
providing, respectively, constraints from void extra Higgs boson searches and parameter determinations from the discovered Higgs state. 
The parameters, in the $\kappa$ `coupling modifier' scheme \cite{LHCHiggsCrossSectionWorkingGroup:2012nn} are extrapolated from the present values extracted at about 30 fb$^{-1}$ of integrated luminosity (after Run 1 and into Run 2) to 300 fb$^{-1}$ (end of Run 3) and 3000 fb$^{-1}$ (HL-LHC) by adopting the expected experimental accuracies given in the scenario 2 of Ref.~\cite{CMS:2013xfa}.
Among the  $\kappa$ modifiers, those interesting us are primarily $\kappa_{VV}^h$ ($V=W^\pm,Z$), $\kappa_{\gamma\gamma}^h$ and $\kappa_{gg}^h$, which are, typically, the most constraining ones.

A completely general 2HDM Lagrangian introduces Higgs-mediated Flavour Changing Neutral Currents (FCNCs) at tree level via Higgs boson exchanges.
In order to tame that,  we assume  an alignment (in flavour space) between the Yukawa matrices as has been done in the elementary Aligned 2HDM (A2HDM)~\cite{Pich:2009sp}. 
In this scenario, the coupling of the heavy Higgs $H$ to the top quark is controlled (plus some small corrections induced by the mixing angle $\theta \sim  v^2/f^2$) by
\begin{align}\label{A2HDM}
\zeta_t=\frac{{\bar\zeta}_t-\tan\beta}{1+{\bar\zeta}_t\tan\beta},
\end{align}
where ${\bar\zeta}_t$ (defined in \cite{DeCurtis:2018zvh}) and $\tan\beta$ (given, as usual, by the ratio of the two Higgs vevs) are predicted, and correlated to each other, in terms of the fundamental parameters of the C2HDM.
Being interested in the phenomenology of the $H$ state, we  map the results of our scan in terms of $m_H$ and $\zeta_t$ and we restrict the parameter space to the region $m_{H,A,H^\pm} >2 m_h$. As an example of the correlations between observables mentioned above, we show in the left panel of Fig.~\ref{fig:flavour} the splitting between the CP-odd state and the heavy CP-even and charged Higgs scalars.

Even though we assume a flavour symmetric composite sector, there are modifications to rare flavour transitions in the SM induced by the exchange of the pesudo-Nambu-Goldstone bosons.
The bound from the $B\to X_s \gamma$ process depends on the interplay between the top and the bottom contributions and, in particular, on the relative size of the charged Higgs $H^\pm$ couplings to the top and bottom quarks $\zeta_t$ and $\zeta_b$. In the scenario discussed in \cite{DeCurtis:2018zvh}, in which $\zeta_t = \zeta_b$, the excluded region from $B\to X_s \gamma$ is shown in Fig.~\ref{fig:flavour}, right panel, by the red shading.
 In different scenarios providing $\zeta_b < \zeta_t$ the bound can be relaxed such  that all the points survive the constraint. 
On the other hand, the bound from the measurement of the $B_s \to \mu^+\mu^-$ transition is more robust as it only depends on $\zeta_t$. However, the corresponding excluded region does not overlap with the distribution of phenomenologically accepted points.

\begin{figure}
\begin{center}
{\includegraphics[width=0.45\textwidth]{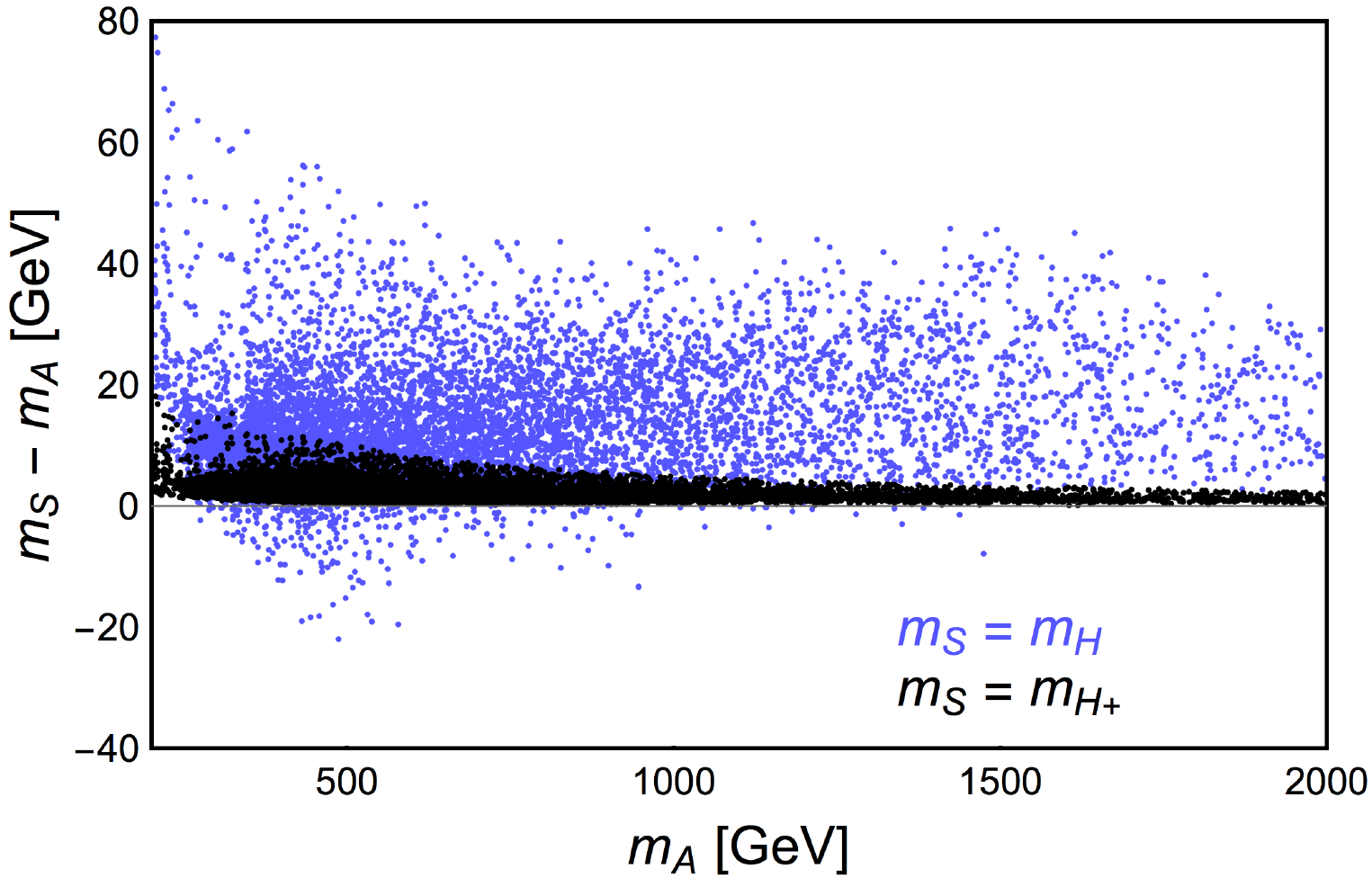}}
{\includegraphics[width=0.45\textwidth]{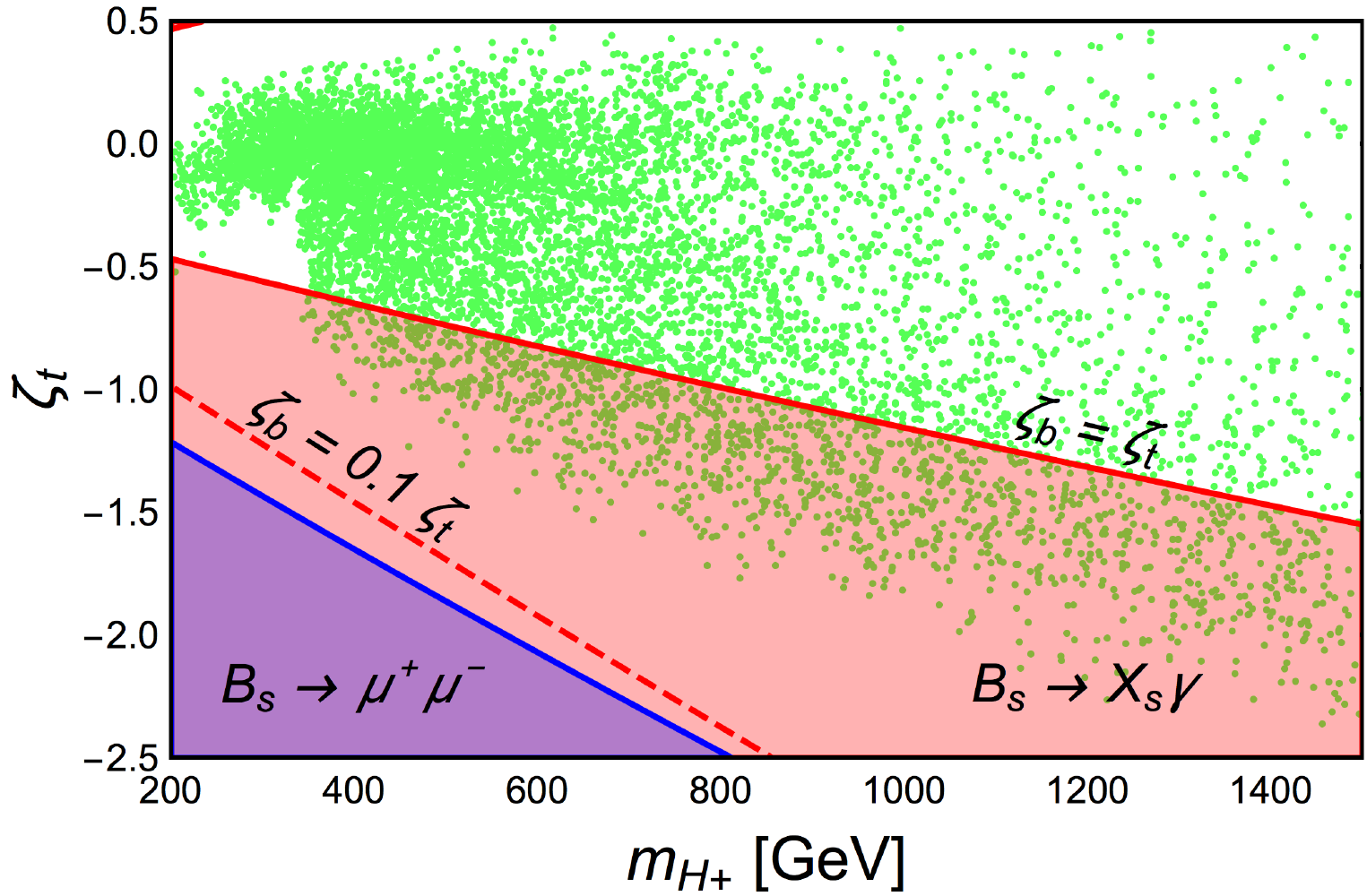}}
\caption{\label{fig:flavour} 
(Left) Mass splittings $m_H - m_A$ (blue) and $m_{H^+} - m_A$ (black) versus $m_A$.
(Right) Correlation between the charged Higgs mass $m_{H^+}$ and the corresponding coupling to the top quark. Green points are allowed by current LHC direct and indirect searches. Red and purple regions are excluded at $2\sigma$ level by measurements of the $B \rightarrow X_s \gamma$ (under the assumption $\zeta_b = \zeta_t$) and of $B_s \rightarrow \mu^+ \mu^-$ transitions, respectively. The former bound strongly depends on the size of $\zeta_b$}
\end{center}
\end{figure}
}

\begin{figure}
\centering
{\includegraphics[scale=0.45]{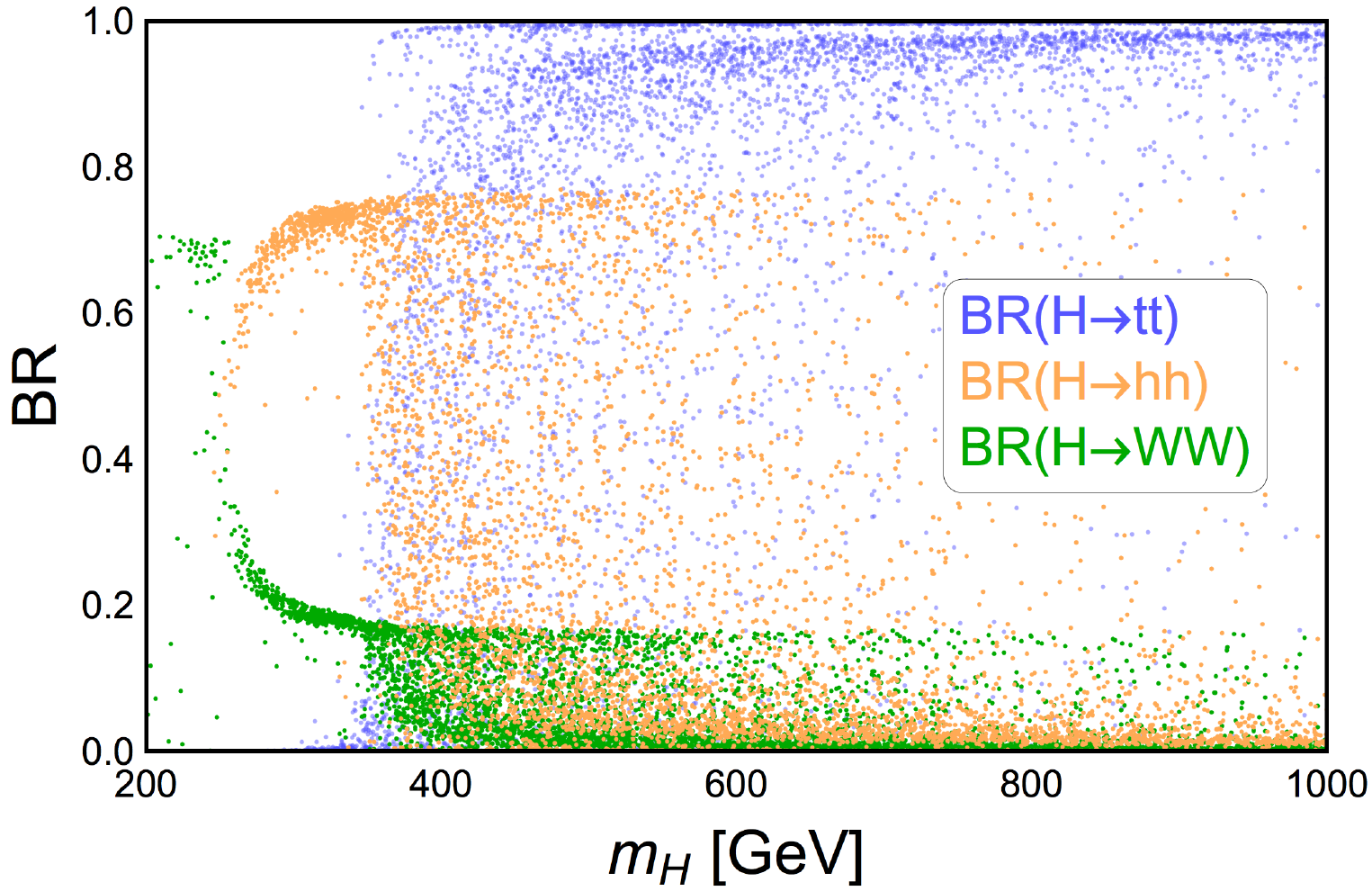}} \quad
{\includegraphics[scale=0.45]{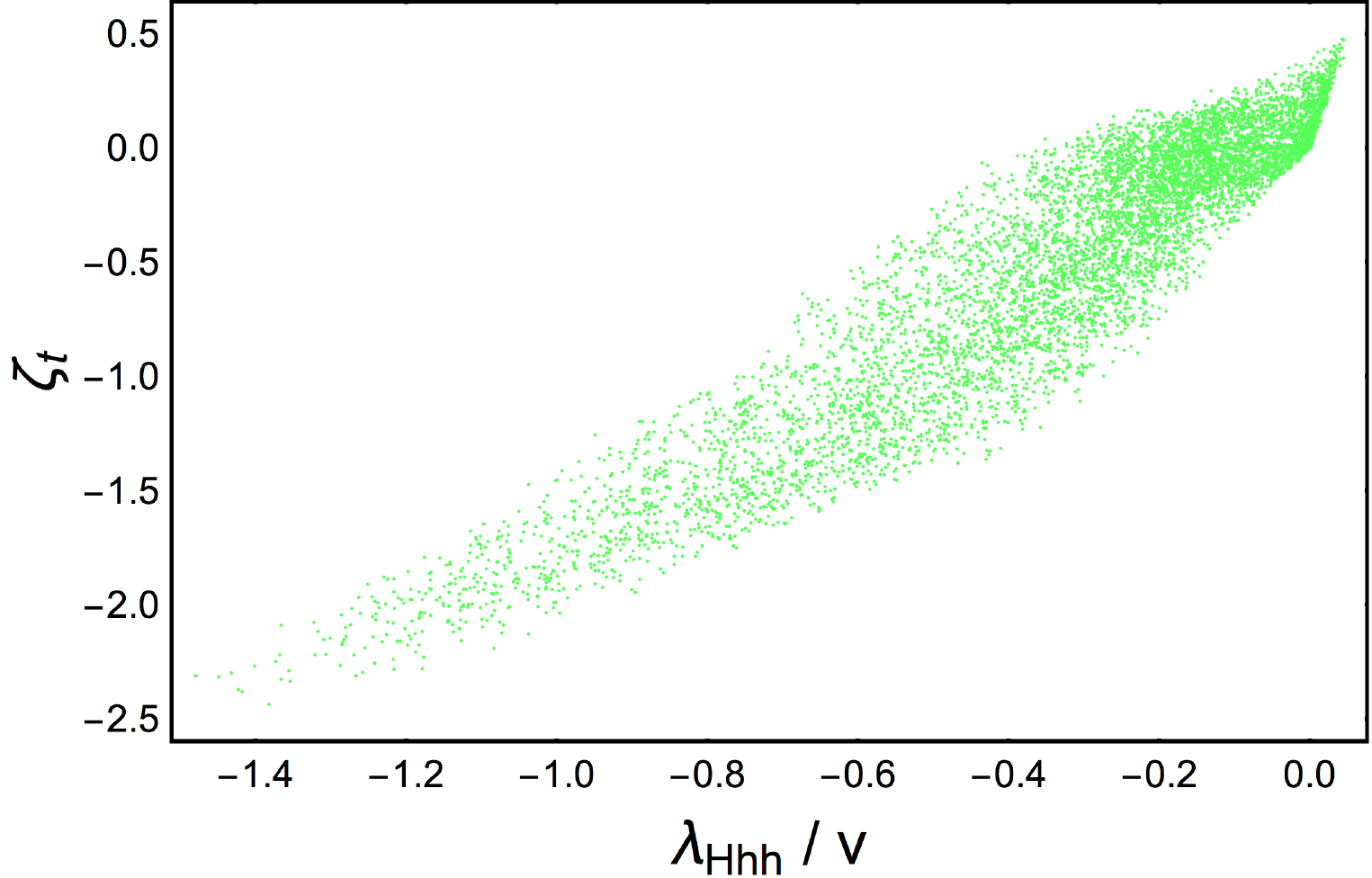}}
\caption{(Left) The BR of the $H$ boson of the C2HDM as a function of its mass in the following decay channels: $WW$ (green), $hh$ (orange) and $t\bar t$ (blue). (Right) the correlation between the couplings $\zeta_t$ and $\lambda_{Hhh}$ obtained upon imposing current HiggsBounds and HiggsSignals constraints at 13 TeV. 
\label{fig:BRs}}
\end{figure}

The parameter $\zeta_t$ and the Higgs trilinear coupling $\lambda_{Hhh}$ control the hierarchy among the decay channel of the heavy $H$ state. 
In particular, $H \rightarrow t \bar t$, when kinematically allowed, represents the main decay mode. Below the $t \bar t$ kinematical threshold, the di-Higgs $H \rightarrow hh$ mode can reach $\sim 80\%$ while the remaining decay space is saturated by $H \rightarrow VV$. The corresponding Branching Ratios (BRs)  are shown in the left panel of Fig.~\ref{fig:BRs} and both of these can be much different in the C2HDM with respect to elementary realisations of the 2HDM, since the $Hhh$ and $Ht\bar t$ interactions carry the imprint of compositeness (their correlation is shown in the plot on the right of Fig.~\ref{fig:BRs}).
The hierarchy between the $H \rightarrow hh$ and $H \rightarrow t \bar t$ decay modes highlights their key role in the discovery and characterisation of the composite heavy Higgs boson.

\begin{figure}
\centering
{\includegraphics[scale=0.45]{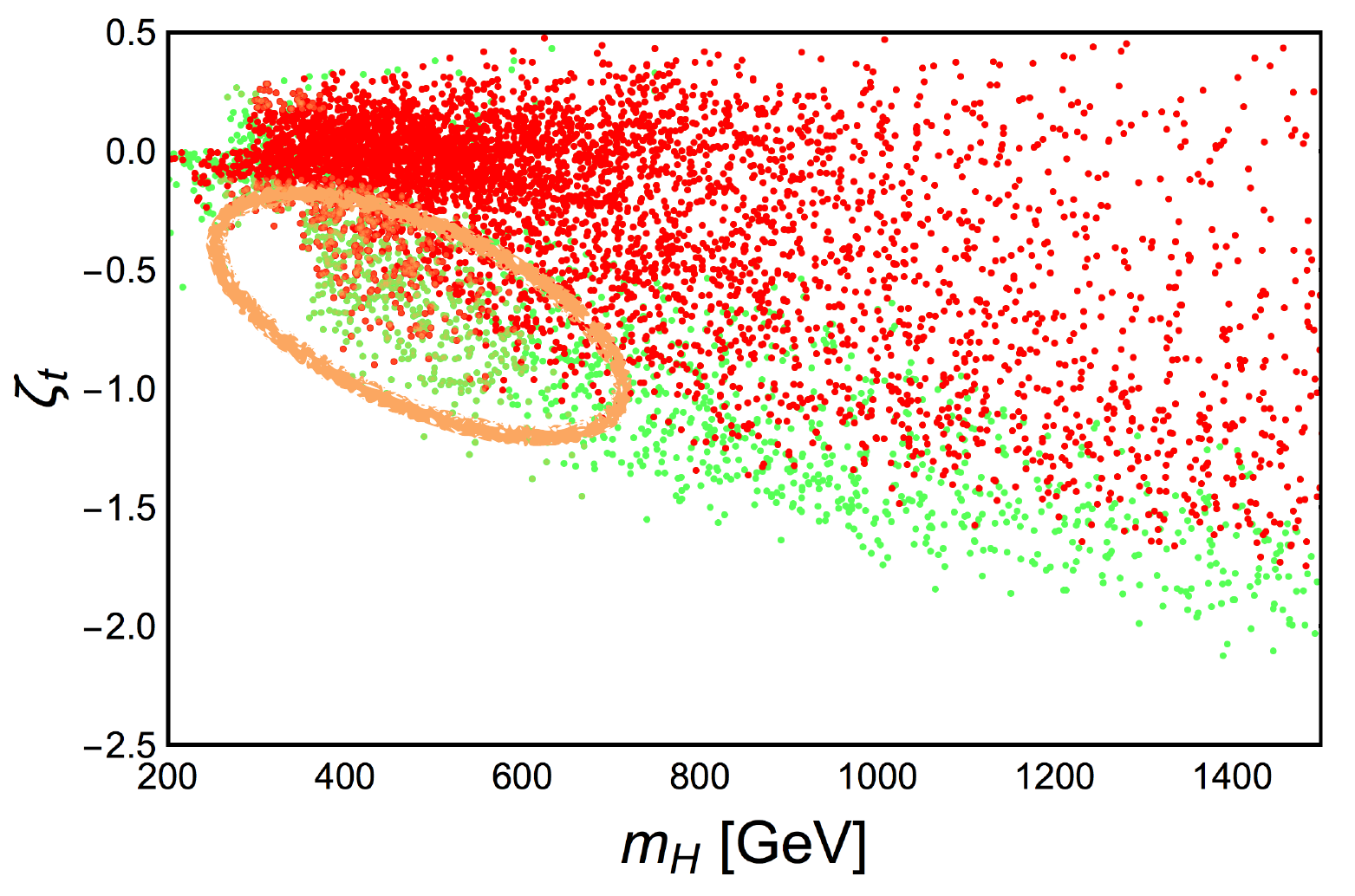}} \qquad
{\includegraphics[scale=0.45]{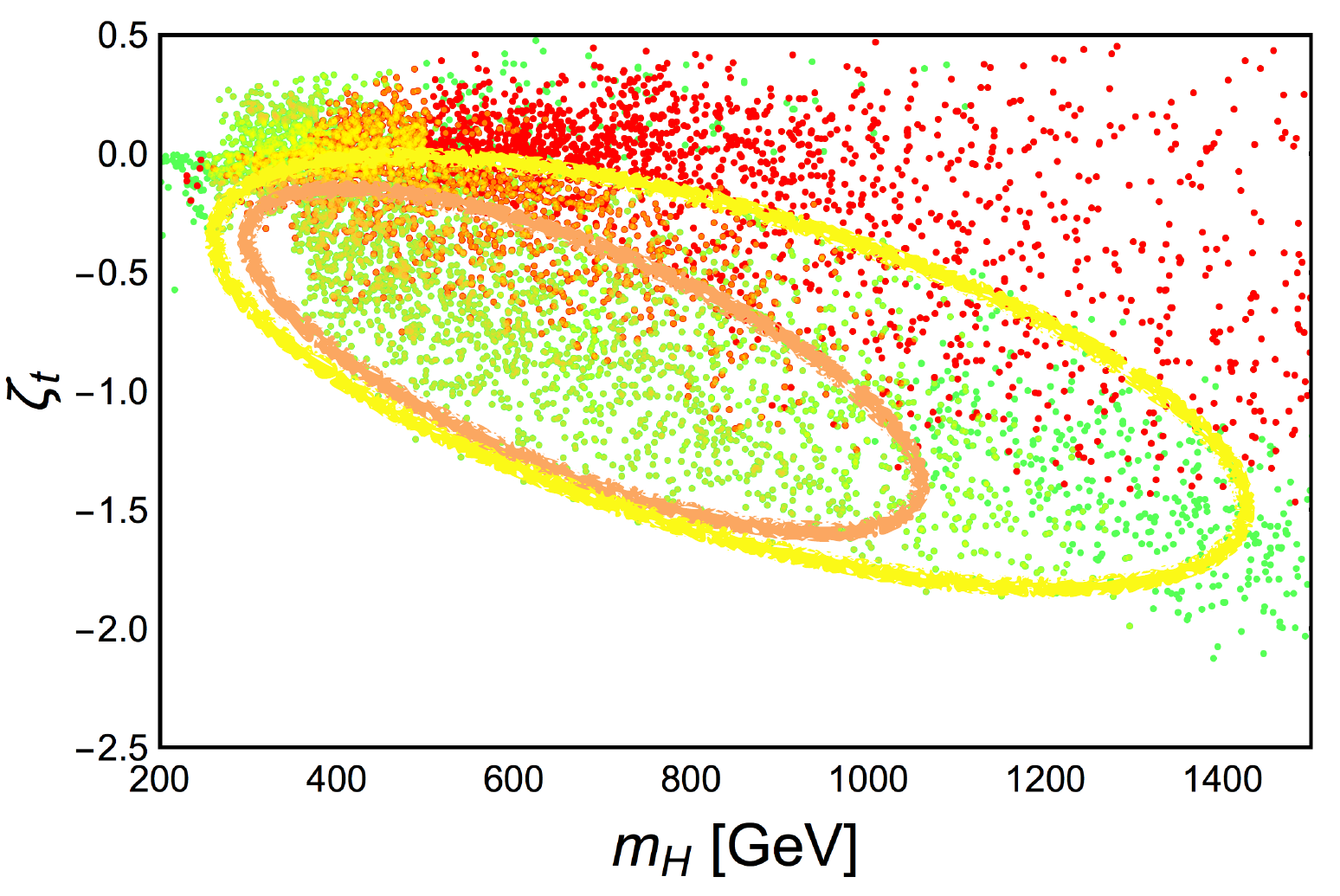}}
\caption{Results of the C2HDM scan described in the text. Colour coding is as follows.
Green: all points that pass present constraints at 13 TeV. 
Red: points that, in addition to the above, have $\kappa_{VV}^h$, $\kappa_{\gamma\gamma}^h$ and $\kappa_{gg}^h$ within the 95\% CL projected uncertainty at $L=300$ fb$^{-1}$ (left) and $L=3000$ fb$^{-1}$ (right).
Orange: points that, in addition to the above,  are 95\% CL excluded  by the direct search $gg\to H\to hh\to b\bar b\gamma\gamma$, at $L=300$ fb$^{-1}$ (left) and $L=3000$ fb$^{-1}$ (right). In the right plot the yellow points are 95\% CL excluded by the same search  at the HE-LHC with $\sqrt{s} = 27$ TeV and $L=15$ ab$^{-1}$. The orange and yellow elliptical shapes highlight the regions in which the points of the corresponding colour accumulate.
\label{fig:bbgaga}}
\end{figure}

Figs.~\ref{fig:bbgaga} show the interplay between direct and indirect searches and the capabilities of the HL-LHC and HE-LHC in the discovering of the $gg\to H\to hh\to b\bar b\gamma\gamma$  signal over regions of the C2HDM parameter space projected onto the plane $(m_H,\zeta_t)$, even when no deviations are visible in the $\kappa$ modifiers of the SM-like Higgs state $h$ (red points) with luminosities $L=300$ fb$^{-1}$ and $L=3000$ fb$^{-1}$. 
The 95\% Confidence Level (CL) exclusion limits have been extracted by employing the sensitivity projections declared in \cite{Aaboud:2018mjh} and 
\cite{CMS:2017ihs} while compliance with the coupling modifiers has been achieved by asking that $|1 - k^h_i|$ is less than the uncertainty discussed in Ref.~\cite{CMS:2013xfa}, with $i=VV,\gamma\gamma$ and $gg$. 
Notice that  the orange points have a large overlap with the red ones for small values of $| \zeta_t| $. 
As shown in \cite{CidVidal:2018eel}, the $gg\to H\to t \bar t$ (followed by semi-leptonic decays) would also allow to probe larger values of $m_H$.
Therefore, the $gg\to H\to hh$ process enables one to cover a larger C2HDM parameter space while the $gg\to H\to t \bar t$ one higher $H$ masses. 
As such, the combination of the two allows to obtain the benefits of either.  
The  HE-LHC,  assuming $\sqrt s=27$ TeV and $ L=15$ ab$^{-1}$,  will improve the reach in the $H$ high mass region up to 1.3 TeV by studying the process $gg\to H\to hh\to b\bar b\gamma\gamma$ (see the right plot in Fig.~\ref{fig:bbgaga}). Concerning the $gg \to H\to t\bar t$  channel,  the naive extrapolation of the sensitivity with the parton luminosities at the HE-LHC is unreliable because it is affected by the SM $t \bar t$ threshold effects. We also remark that for a proper phenomenological analysis of the $t\bar t$ process, the interference effects with $gg$-induced irreducible background must be fully taken into account. 
Indeed, the interference effects between the $gg$-induced QCD diagrams at leading order and the one providing a Higgs boson in $s$-channel via gluon fusion generate a peak-dip structure of the $M_{t\bar t}$ spectrum
that could sensibly affect the sensitivity reach of this process. 

In summary, in view of the phenomenological results produced for the composite realisation of the 2HDM in presence of the up-to-date experimental constraints, we are confident to have highlighted the main distinctive features between  this scenario and the elementary  one to be tested at the LHC by the end of all its, already scheduled and currently discussed, future stages.

\section*{Acknowledgements}
SM is supported in part through the NExT Institute and the STFC CG ST/L000296/1. SM and KY acknowledge financial support from the H2020-MSCA-RISE-2014 grant no. 645722 (NonMinimalHiggs).

\providecommand{\href}[2]{#2}\begingroup\raggedright

\end{document}